\documentclass[lettersize,journal]{IEEEtran}

\PassOptionsToPackage{hyphens}{url}
\usepackage{setspace}

\usepackage{amsmath}
\usepackage{amsthm}
\usepackage{amssymb}    

\usepackage{makecell}
\usepackage{longtable}
\usepackage{booktabs}
\usepackage{float}
\usepackage[table]{xcolor}   
\usepackage{multirow}
\usepackage{stfloats}

\usepackage[ruled,vlined]{algorithm2e}  

\usepackage{graphicx}
\usepackage{subfig}
\usepackage{url}
\usepackage{cite}
\usepackage{textcomp}

\usepackage{flushend}
\usepackage[colorlinks,linkcolor=red,anchorcolor=green,citecolor=blue]{hyperref}
\usepackage{cleveref}

\usepackage{bbding}
\usepackage{verbatim}
\crefname{figure}{fig}{figures}
\Crefname{figure}{Fig}{Figures}
\begin{document}

\title{A Novel Hierarchical Co-Optimization Framework for Coordinated Task Scheduling and Power Dispatch in Computing Power Networks}

\author{Haoxiang Luo, Kun Yang, Qi Huang,~\IEEEmembership{Fellow,~IEEE}, Marco Aiello,~\IEEEmembership{Senior Member,~IEEE}, \\Schahram Dustdar,~\IEEEmembership{Fellow,~IEEE}

\thanks{H. Luo is with the School of Information and Communication Engineering, University of Electronic Science and Technology of China, Chengdu 611731, China (e-mail: lhx991115@163.comn). K. Yang is with the Leshan Power Supply Company, State Grid Sichuan Electric Power Company, Leshan 613100, China (e-mail: 112676539@qq.com). Q. Huang is with the School of Mechanical and Electrical Engineering, University of Electronic Science and Technology of China, Chengdu 611731, China, and also with the Institute of Scientific and Technical Information of China, Beijing 100038, China (e-mail: hwong@uestc.edu.cn). M. Aiello is with the Service Computing Department, Institute for Architecture of Application Systems, University of Stuttgart, 70174 Stuttgart, Germany (e-mail: marco.aiello@iaas.uni-stuttgart.de). S. Dustdar is with the ICREA, Barcelona 08002, Spain (e-mail: schahram.dustdar@upf.edu).}

}

\maketitle

\begin{abstract}
The proliferation of large-scale AI and data-intensive applications has driven the development of Computing Power Networks (CPN). It is a key paradigm for delivering ubiquitous, on-demand computational services with high efficiency. However, CPNs face dual challenges in service computing. Immense energy consumption threatens sustainable operations.  And the integration with power grids also features high penetration of intermittent Renewable Energy Sources (RES), complicating task scheduling while ensuring Quality of Service (QoS). To address these issues, this paper proposes a novel Two-Stage Co-Optimization (TSCO) framework. It synergistically coordinates CPN task scheduling and power system dispatch, aiming to optimize service performance while achieving low-carbon operations. The framework decomposes the complex, large-scale problem into a day-ahead stochastic unit commitment stage and a real-time operational stage. The former is solved using Benders decomposition for computational tractability, while in the latter, economic dispatch of generation assets is coupled with an adaptive CPN task scheduling managed by a deep reinforcement learning agent. It makes carbon-aware decisions by responding to dynamic grid conditions, including real-time electricity prices and marginal carbon intensity. Extensive simulations demonstrate that the TSCO outperforms baseline approaches significantly. It reduces carbon emissions by $16.2\%$ and operational costs by $12.7\%$, while decreasing RES curtailment by over $60\%$, maintaining a task success rate of $98.5\%$, and minimizing average task tardiness to $12.3$s. This work advances cross-domain service optimization in CPNs.

\end{abstract}


\section{Introduction} \label{sec-I}
\subsection{Background}
\IEEEPARstart {T}{he} digital transformation of the global economy is fueling an unprecedented demand for computational power. The rise of artificial intelligence (AI) has pushed traditional centralized cloud computing architectures to their limits \cite{luo2025toward}. In response, a new paradigm known as the Computing Power Network (CPN) has emerged \cite{yukun2024computing}.   It aims to interconnect vast, distributed, and heterogeneous computing resources into a unified, programmable fabric \cite{liu2025joint}.   The core vision of CPN is to break down resource silos and enable the flexible, on-demand scheduling and allocation of computing, storage, and network resources, thereby enabling novel, low-latency applications. Unlike grid computing or conventional data centers, CPNs explicitly integrate computing task scheduling with power system dispatch.

Concurrently, the global energy sector is undergoing a profound transition towards sustainability, marked by massive investments in Renewable Energy Sources (RES) and Battery Energy Storage Systems (BESS) \cite{impram2020challenges}. This shift presents a symbiotic relationship with the burgeoning CPN paradigm. On the one hand, CPNs are voracious consumers of electricity; data centers alone are projected to account for a significant portion of global electricity load growth, with AI workloads being a primary driver. Globally, data centers already account for $1-2\%$ of total electricity consumption, a figure comparable to the aviation industry. Driven by the explosive growth of AI, this demand is projected to double by 2030 \cite{chen2025data}. In the United States, data center electricity usage is forecast to climb from $4.4\%$ of the national total in 2023 to as high as $12\%$ by 2028\footnote{https://www.energy.gov/articles/doe-releases-new-report-evaluating-increase-electricity-demand-data-centers}, with AI workloads becoming a primary driver of this expansion. The availability of clean, renewable energy offers a direct path to decarbonize this massive computational infrastructure\cite{teng2024privacy}. On the other hand, the very nature of CPNs, with their inherent flexibility in workload scheduling, presents a unique opportunity to support and stabilize a grid increasingly reliant on intermittent RES \cite{wang2025providing}.

\subsection{Research Motivation}

The core challenge lies in a misalignment between the operational dynamics of CPNs and RES-dominated power grids. CPNs are designed to host a wide array of computational tasks, from latency-sensitive services, to compute-intensive, batch-processing jobs \cite{luo2024bc4llm}. These tasks demand a highly reliable and stable power supply to ensure Quality of Service (QoS). However, RES are inherently variable and dependent on weather conditions. This intermittency introduces significant volatility into the power grid. It often results in the wasteful curtailment of clean energy when generation exceeds demand \cite{jiang2025bargaining}.  

Simply connecting a CPN to a grid with high RES penetration without intelligent coordination creates a direct conflict. During periods of low RES output, the CPN's demand would force the grid to rely on expensive and carbon-intensive fossil-fuel peaker plants to maintain balance. Conversely, during periods of high RES output, the grid may be forced to curtail wind or solar generation to prevent overload, even as CPNs continue to draw power from a mixed-carbon source. This decoupled operation leads to a suboptimal outcome: either unreliable computation or high operational costs and carbon emissions. The problem is not merely about sourcing green energy, but about managing the flexibility of CPN workloads as a grid-stabilizing asset. The inherent ability to shift CPN tasks in time (delaying non-critical jobs) and space (migrating workloads to regions with abundant RES) constitutes a powerful form of Demand Response (DR) \cite{yang2024secure}, \cite{piontek2023carbon}. This reframes the problem from ``\emph{how to power the CPN cleanly}" to ``\emph{how to leverage the CPN's flexibility to enable a cleaner, more stable grid}". This symbiotic relationship is the core motivation for the co-optimization framework proposed in this paper.

\subsection{Contributions of the Paper}

In the present work, we propose a holistic, hierarchical co-optimization framework that jointly optimizes power system operations and CPN task scheduling. By breaking down the silos between power system operators and CPN schedulers, the framework unlocks significant economic and environmental benefits. It coordinates power generation dispatch with computational task scheduling across multiple timescales to align CPN energy demand with the availability of low-cost, low-carbon renewable energy. To the best of our knowledge, this is one of the first efforts to jointly optimize the power grid and the CPN. The main contributions of this work are summarized next:

\begin{itemize} 

\item \textbf{A Comprehensive Integrated System Model:} A detailed model is developed that captures the intricate interplay between a heterogeneous CPN and a modern power grid. It incorporates conventional thermal generators, stochastic RES, and BESS, while explicitly modeling the spatio-temporal dynamics of electricity prices and carbon intensity. 

\item \textbf{A Two-Stage Co-Optimization (TSCO) Framework:} A hierarchical framework is designed to decompose the computationally intractable joint optimization problem into two manageable stages. A day-ahead planning stage addresses long-term unit commitment and resource reservation, while a real-time operational stage handles dynamic economic dispatch and adaptive task scheduling.

\item \textbf{Scalable Optimization with Benders Decomposition:} To address the large-scale, mixed-integer nature of the day-ahead Stochastic Unit Commitment (SUC) problem, Benders decomposition is employed. This technique effectively decouples the integer commitment decisions from the continuous dispatch variables, ensuring the problem remains computationally tractable even for large systems and numerous uncertainty scenarios. 

\item \textbf{Adaptive Real-Time Scheduling with Deep Reinforcement Learning (DRL):} A DRL agent is developed for the real-time CPN task scheduling. This model-free approach enables fast, adaptive, and carbon-aware scheduling decisions in response to the highly dynamic and complex state of the joint CPN-grid system.


\end{itemize}

\subsection{Paper Structure}

The remainder of the paper is structured as follows. Section~\ref{sec-II} provides a review of related work in CPN scheduling, integrated energy systems, and carbon-aware computing. Section \ref{sec-III} presents the detailed mathematical formulation of the CPN and power system models, defining the joint optimization problem. Section \ref{sec-Iv} describes the proposed TSCO framework, including the Benders decomposition algorithm and the DRL-based scheduler. Section \ref{sec-v} details the simulation setup and presents a comprehensive performance evaluation against several baseline methods. Finally, Section \ref{sec-vi} concludes the paper with a summary of findings and directions for future research.

\section{Related Works}\label{sec-II}

\subsection{Computing Power Network Architectures and Scheduling}

The CPN concept has evolved from earlier paradigms, such as cloud, fog, and edge computing, with the primary goal of creating a unified network for ubiquitous computing resources. Early research focused on defining the architecture and core features of CPNs, such as intent-driven operation, closed-loop autonomy, and elastic scheduling. Architectures have been proposed with both centralized control planes, which possess a global view for unified scheduling, and distributed schemes, where decisions are made locally by network nodes. 

A significant body of research in CPNs has concentrated on task scheduling \cite{xie2025priority}. The primary objectives have traditionally been to optimize QoS metrics. For instance, studies have focused on developing scheduling policies to minimize task completion delay and enhance reliability, often formulating the problem as a Continuous-Time Markov Decision Process (CMDP) and solving it with DRL techniques \cite{ma2025optimizing}. Other works have explored task offloading in terminal-side CPNs or the joint selection of routing paths and computing nodes \cite{chen2024two}. 

Two other works also considered the collaboration between the CPN and the power grid. Wen et al. \cite{wen2026green}, \cite{wen2024spatiotemporal} proposed a spatiotemporal task scheduling scheme for CPNs, leveraging a customized DRL algorithm to optimize energy consumption and carbon emissions within CPNs. However, their work focuses solely on intra-domain optimization, treating energy supply as an exogenous input without involving power system dynamics. Zhong et al. \cite{zhong2024joint} integrated Electric Vehicles (EVs) into CPN and Power Distribution Network (PDN) coordination. They use EVs as dual carriers of energy storage and computing resources for task offloading. Nevertheless, their CPN scheduling is subordinate to cost minimization of EV-PDN synergy, lacking deep coupling with bulk power systems and carbon-aware optimization. In contrast, our TSCO framework emphasizes bidirectional coupling between CPN and bulk power systems, treating CPN as a flexible load to enhance grid stability and low-carbon operations.
Table \ref{tab:cpn_scheduling_comparison} compares the approaches with ours.

While some research considers energy consumption as a constraint or a secondary objective, the explicit, primary optimization of carbon emissions based on the real-time state of the power grid remains largely unexplored. 

\begin{table*}[!t]
\centering
\caption{Comparison of CPN and Power Grid Coordination}
\label{tab:cpn_scheduling_comparison}
\setlength{\tabcolsep}{3pt} 
\fontsize{8pt}{9.6pt}\selectfont 
\begin{tabular}{m{2.8cm}m{4.7cm}m{4.7cm}m{4.7cm}} 
\toprule
\textbf{Key Dimension} & \textbf{TSCO (Our Work)} & \textbf{\cite{wen2026green}, \cite{wen2024spatiotemporal}} & \textbf{\cite{zhong2024joint}} \\
\midrule
Research Paradigm & Cross-domain co-optimization & Intra-CPN fine-grained scheduling & Multi-agent synergy \\

Collaborative Entities & CPN, bulk power system & CPN & CPN, PDN, EVs\\

Optimization Core & Low carbon, RES curtailment reduction, grid stability, communication QoS & Energy efficiency, task delay, load balancing & System cost minimization\\

System Interaction & Bidirectional (CPN and grid) & No interaction& Unidirectional (CPN/EV constrained by PDN) \\

Key Innovation & Bidirectional coupling; endogenous carbon intensity & Triple-selection (nodes/paths/time); intra-CPN coordination & EV dual-resource (energy/computing) exploitation\\

\bottomrule
\end{tabular}
\vspace{-0.5cm}
\end{table*}

\subsection{Integrated Energy System Management}

In the power systems domain, the concept of coordinating generation, transmission, and consumption has been studied extensively under the ``Generation-Grid-Load-Storage" integrated operation model \cite{luo2024multi}. This paradigm seeks to improve system safety, economy, and reliability through the coordinated interaction of all components. Research in this area includes the development of multi-timescale optimal dispatching strategies \cite{ma2024study}, economic dispatch models that incorporate DR, and Optimal Power Flow (OPF) formulations that aim to minimize generation costs while respecting network constraints \cite{chowdhury2025optimal}.

The integration of flexible loads and DR has been identified as a key enabler for grids with high RES penetration. Studies have explored how to coordinate data centers as flexible loads with a load aggregator to minimize electricity costs and absorb grid volatility \cite{zhang2025unlocking}. However, these studies often rely on simplified models of the flexible load, such as an abstract ability to shift power consumption in time, without capturing the complex internal constraints, dependencies, and heterogeneous resource requirements of a CPN workload. They treat the CPN as a ``closed box" load, missing the opportunity to optimize its internal operations in concert with the grid.

\subsection{Carbon-Aware Workload Scheduling}

A substantial body of work on carbon-aware scheduling for geographically distributed data centers. A common strategy is spatio-temporal scheduling \cite{ye2024deep}, which involves shifting computational workloads in time or space to data centers with lower electricity prices or cleaner energy mixes. These methods often leverage real-time carbon intensity to guide scheduling.

To manage carbon emissions over the long term, some works have proposed online algorithms based on Lyapunov optimization \cite{ma2025greening}. This technique transforms a long-term average constraint into a series of real-time optimization subproblems by maintaining a ``virtual queue" that tracks the deviation from the budget. The scheduler is then penalized for actions that increase this queue length \cite{xu2024optimal}. This line of research suffers from a critical limitation.  It almost universally treats the power grid as an exogenous system. The price and carbon intensity signals are assumed to be external inputs that are unaffected by the scheduling decisions. This assumption breaks down at scale, as the collective actions of large CPNs can influence grid operations, market prices, and the generation mix. Specifically, we have done a study comparing four countries: Germany, the Netherlands, France, and Brazil, as these have very different energy mixes and therefore the variability of the carbon intensity is very different among them.


\subsection{Research Gap Summary}

The existing body of work, while extensive in its respective domains, reveals a significant research gap at the intersection of CPNs and power systems. Current research either:
\begin{itemize}
    \item \textbf{Simplifies the power grid:} CPN and carbon-aware scheduling studies treat the grid as a static source of price and carbon signals, ignoring the feedback loop where scheduling decisions impact the grid.
    \item \textbf{Simplifies the CPN:} Power system and DR studies model flexible loads in an overly simplistic manner, failing to capture the rich internal complexity of CPN workloads, resource heterogeneity, and QoS constraints.
\end{itemize}

This leads to a new class of control problems where the decision variables are distributed across two separate domains. The system dynamics are characterized by a mix of well-understood physics (the power grid) and complex, stochastic behavior (the CPN). A model-based approach is ill-suited for fast, dynamic CPN scheduling, while a model-free AI approach cannot guarantee adherence to hard physical constraints of the power grid. Consequently, there is a need for a novel framework that endogenously models the bidirectional interactions between them. Our TSCO framework, which combines large-scale, model-based optimization for slow, physics-heavy planning with a model-free, adaptive AI technique for fast, complex real-time scheduling, is designed specifically.

\section{System Architecture and Problem Formulation}\label{sec-III}
\begin{figure}[!t]
   \centering
   \includegraphics[width=3 in]{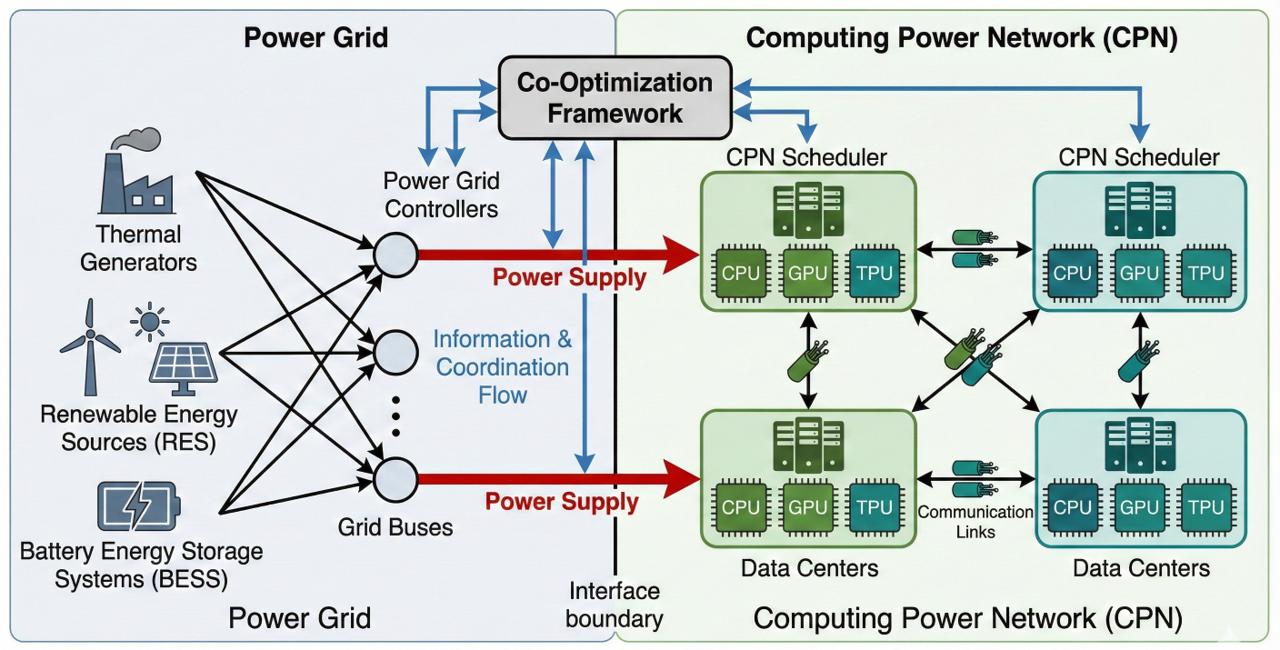}
   \caption{CPN and power grid co-optimization architecture.}
   \label{fig0}
   \vspace{-0.5cm}
\end{figure}

These models form the basis of our co-optimization framework, as shown in Fig. \ref{fig0}. 

\subsection{Computing Power Network Model}

The CPN is modeled as a directed graph $\mathcal{G}_{CPN} = (\mathcal{N}, \mathcal{L})$, where $\mathcal{N}$ is the set of geographically distributed CPN nodes, and $\mathcal{L}$ is the set of communication links connecting them.

\subsubsection{CPN Node Model} Each CPN node \(n \in N\) consists of heterogeneous hardware resources, denoted by the set \(H = \{CPU, GPU, TPU\}\). For each hardware type \(h \in H\), we define its key parameters:  \(C_{n,h}^{comp}\) is computing capacity in FLOPS); \(P_{n,h}^{idle}\) is idle power consumption; \(P_{n,h}^{peak}\) denotes peak power consumption at full utilization; \(\alpha_{n,h}, \beta_{n,h}\) are nonlinear coefficients calibrated from real-world hardware traces.

The power consumption of hardware $h$ at node $n$ follows a quadratic nonlinear model, capturing the superlinear power growth of accelerators (e.g., GPU/TPU) under high utilization:
\begin{equation}
P_{n,h,t}^{comp} = P_{n,h}^{idle} + (\alpha_{n,h} \cdot u_{n,h,t}^2 + \beta_{n,h} \cdot u_{n,h,t}) \cdot (P_{n,h}^{peak} - P_{n,h}^{idle}),
\end{equation}
where \(u_{n,h,t}\) is the utilization of hardwar $h$ at node $n$ and time $t$, calculated as:
\begin{equation}
u_{n,h,t} = \frac{\sum_{k \in \mathcal{J}} \sum_{\tau \in \mathcal{T}_k} r_{\tau,h} \cdot x_{k,\tau,n,h,t}}{C_{n,h}^{comp}}, 
\end{equation}
where \(r_{\tau,h}\) denotes the resource requirement of subtask \(\tau\) for hardware$h$, set to 0 if \(\tau\) does not require $h$. \(x_{k,\tau,n,h,t}\) is a binary variable. Specifically, $1$ if subtask \(\tau\) of job $k$ is assigned to hardware $h$ at node $n$ and time $t$, $0$ otherwise. The total power consumption of node$n$ and time $t$ is the sum of power from all hardware types:
\begin{equation} 
P_{n,t}^{CPN} = \sum_{h \in H} P_{n,h,t}^{comp}.
\end{equation}

\subsubsection{Task Model} We model incoming computational jobs as Directed Acyclic Graphs (DAGs), a common representation for parallel applications with precedence constraints \cite{chen2024drdst}, \cite{xie2017energy}. A job $k \in \mathcal{J}$ is represented by $J_k = (\mathcal{T}_k, \mathcal{E}_k)$, where $\mathcal{T}_k$ is the set of sub-tasks and $\mathcal{E}_k$ is the set of directed edges representing dependencies. An edge $(\tau_i, \tau_j) \in \mathcal{E}_k$ implies that sub-task $\tau_j$ cannot begin until $\tau_i$ is complete. Each sub-task $\tau \in \mathcal{T}_k$ is defined by its total computational workload $w_{\tau}$ (in floating-point operations) and its resource requirement $r_{\tau}$ (e.g., number of processing units). The execution time of sub-task $\tau$ on node $n$ is thus $t_{\tau,n} = w_{\tau} / C_n^{comp}$. Each job $k$ has an arrival time $A_k$ and a hard end-to-end deadline $D_k$.

Each subtask \(\tau \in T_k\)  is further defined by: (1) hardware type constraint \(h_\tau \subseteq H\),  the set of hardware types capable of executing \(\tau\), e.g., large-model training subtasks require \(h_\tau = \{GPU, TPU\}\); (2) data volume \(d_\tau\) in GB, for communication constraint calculation; (3) total computational workload \(w_\tau\) in FLOPs, and resource requirement \(r_{\tau,h}\) is the number of processing units for hardware \(h \in h_\tau\). The execution time of subtask \(\tau\) on hardware $h$ of node $n$ is:

\begin{equation} 
t_{\tau,n,h} = 
\begin{cases} 
\frac{w_\tau}{C_{n,h}^{\mathrm{comp}}}, & \text{if } h \in h_\tau \\
\infty, & \text{otherwise}
\end{cases}
\quad \forall n, h \in h_\tau, \tau \in \mathcal{T}_k.
\end{equation} 
\subsubsection{Communication Link Model} The communication network between CPN nodes is modeled as a directed graph \(\mathcal{G}_{comm} = (\mathcal{N}, \mathcal{L})\), where $\mathcal{L}$ denotes the set of bidirectional communication links between nodes. For each link from the source node to the destination node \((n_{src}, n_{dest}) \in \mathcal{L}\), two key parameters are defined based on real-world network traces:
\begin{itemize}

    \item \textbf{Bandwidth:} \(B_{n_{src},n_{dest}}\) (GB/s) is the maximum data transmission rate of the link, fixed for dedicated CPN networks.
     \item \textbf{Latency parameters:} It includes \(lat_{n_{src},n_{dest}}^{base}\) (s, base latency without data transmission) and \(\gamma_{n_{src},n_{dest}}\) (s/GB, dynamic latency coefficient related to data volume).
Then, the total latency for transmitting a subtask \(\tau\) with data volume \(d_\tau\) from \(n_{src}\) to \(n_{dest}\) is:
\begin{equation}
 lat_{n_{src},n_{dest}}(\tau) = lat_{n_{src},n_{dest}}^{base} + \gamma_{n_{src},n_{dest}} \cdot d_\tau. 
 \end{equation}

The data migration time (i.e., the duration of data transmission) is determined by link bandwidth:
\begin{equation}
\label{eq6}
t_{migrate,\tau,n_{src},n_{dest}} = \frac{d_\tau}{B_{n_{src},n_{dest}}}. 
\end{equation}

\end{itemize}

\subsection{Integrated Power System Model}

 The system consists of a set of buses $\mathcal{I}$ connected by transmission lines. Each CPN node $n \in \mathcal{N}$ is co-located with a specific load bus $i \in \mathcal{I}$.

\subsubsection{Conventional Generation}The set of conventional thermal generators, $\mathcal{G}_C$, forms the dispatchable backbone of the system. The fuel cost of each generator $g \in \mathcal{G}_C$ is represented by a quadratic function of its power output $P_{g,t}$:
\begin{equation}
C_g(P_{g,t}) = a_g P_{g,t}^2 + b_g P_{g,t} + c_g,
\end{equation}
where $a_g, b_g, c_g$ are cost coefficients. These generators are subject to operational constraints, including minimum and maximum power output limits ($P_g^{min}, P_g^{max}$), and ramp-up/ramp-down rate limits ($RU_g, RD_g$) that constrain how quickly their output can change between time periods \cite{rengel2023optimal}.

\subsubsection{RES}
The set of RES generators, $\mathcal{G}_R$, includes wind and solar farms. Their power output is non-dispatchable and uncertain. We model their available power at time $t$ in scenario $\omega$, $P_{g,t,\omega}^{R,avail}$, as a stochastic parameter derived from historical weather data. The actual dispatched power $P_{g,t,\omega}$ can be less than or equal to the available power.

\subsubsection{BESS}
BESS units, located at specific buses, provide crucial flexibility for managing RES intermittency. Each BESS $b \in \mathcal{B}$ is modeled by its state-of-charge (SOC) dynamics \cite{maroufi2025power}:
\begin{equation}
SOC_{b,t,\omega} = SOC_{b,t-1,\omega} + (\eta_b^c P_{b,t,\omega}^{chg} - \frac{1}{\eta_b^d} P_{b,t,\omega}^{dis}) \Delta t.
\end{equation}
The model is subject to constraints on the SOC level ($SOC_b^{min} \le SOC_{b,t,\omega} \le SOC_b^{max}$), where $E_b^{max}$ is the energy capacity, and maximum charging/discharging power ($P_b^{c,max}, P_b^{d,max}$).

\subsection{Uncertainty and Carbon Modeling}

\subsubsection{Uncertainty Modeling}
The uncertainties in RES generation and CPN job arrivals are critical to the problem. We adopt a scenario-based stochastic programming approach \cite{zou2024aggregator}. A set of discrete scenarios $\Omega$ is generated, where each scenario $\omega \in \Omega$ represents a plausible joint realization of RES power availability and CPN workload over the time horizon $T$. Each scenario is assigned a probability $\pi_\omega$, with $\sum_{\omega \in \Omega} \pi_\omega = 1$.

\subsubsection{Carbon Intensity Modeling}
The environmental impact is quantified through carbon emissions. The carbon intensity is not static but depends on the real-time generation mix. The total carbon emission rate at time $t$ in scenario $\omega$, $E_{t,\omega}$, is calculated as the sum of emissions from all active generators:
\begin{equation}
E_{t,\omega} = \sum_{g \in \mathcal{G}_C} \epsilon_g P_{g,t,\omega},
\end{equation}
where $\epsilon_g$ is the emission factor (e.g., in tons of $CO_2$ per MWh) of generator $g$. For RES, $\epsilon_g$ is zero. This endogenous calculation is crucial, as it directly links dispatch decisions to carbon output. We also leverage real-world marginal carbon intensity data, such as that provided by WattTime\footnote{https://watttime.org/data-science/data-signals/}, which informs the real-time DRL agent about the emission impact resulting from an additional unit of electricity consumption at a specific location and time.

\subsection{Joint Optimization Problem Formulation}
The overarching goal is to co-optimize the power system operation and CPN task scheduling to minimize the total expected system cost over a planning horizon $T$. It comprises the operational costs of the power system and the monetized cost of carbon emissions. It is formulated as a large-scale, two-stage stochastic mixed-integer linear program (MILP).

\textbf{Objective Function:}
\begin{equation}
\min \sum_{\omega \in \Omega} \pi_\omega \sum_{t=1}^{T} \left( \sum_{g \in \mathcal{G}_C} + \lambda_{CO_2} E_{t,\omega} \right).
\end{equation}
The objective minimizes the expected sum of three components across all scenarios: (1) the quadratic fuel costs of conventional generators, (2) the costs associated with starting up ($SU_g$) and shutting down ($SD_g$) these generators based on their commitment status $u_{g,t}$, and (3) a carbon tax, where $\lambda_{CO2}$ is the price of carbon and $E_{t,\omega}$ is the total emissions.

The optimization is subject to a comprehensive set of constraints that couple the two systems:
\begin{itemize}
    \item \textbf{Power System Constraints (for each $t, \omega$):}
   
        \textbf{Power Balance (DC-OPF):} At each bus $i \in \mathcal{I}$, the total power injected must equal the total power withdrawn. This is the core DC power flow equation \cite{bouhouras2024congestion}.
        \begin{multline}
        \sum_{g \in \mathcal{G}(i)} P_{g,t,\omega} + \sum_{b \in \mathcal{B}(i)} (P_{b,t,\omega}^{dis} - P_{b,t,\omega}^{chg}) \\- (P_{i,t}^D + P_{i,t,\omega}^{CPN})  = \sum_{j \in \mathcal{I}} B_{ij}(\theta_{i,t,\omega} - \theta_{j,t,\omega}),
        \end{multline}
        where $\mathcal{G}(i)$ and $\mathcal{B}(i)$ are generators and BESS at bus $i$. $P_{i,t,\omega}^{CPN}$ is the CPN power demand. The right side represents the net power flow out of the bus.
        
         \textbf{Transmission Line Constraints:} The power flow $F_{ij}$ on each line $(i,j)$ must not exceed its thermal limit $F_{ij}^{max}$.
        \begin{equation}
        -F_{ij}^{max} \le B_{ij}(\theta_{i,t,\omega} - \theta_{j,t,\omega}) \le F_{ij}^{max}.
        \end{equation}
         
         \textbf{Generator Constraints:} Including commitment logic, min/max output, and ramping limits for all $g \in \mathcal{G}_C$.
       
        \textbf{BESS Constraints:} Including SOC dynamics, capacity limits, and charge/discharge power limits for all $b \in \mathcal{B}$.
  
    \item \textbf{CPN Task Scheduling Constraints (for each $t, \omega$):}
   
        \textbf{Task Assignment:} Each sub-task $\tau$ of each job $k$ must be scheduled exactly once.
        \begin{equation}
        \sum_{n \in \mathcal{N}}\sum_{h \in \mathcal{h_\tau}}  \sum_{t=A_k}^{D_k} x_{k,\tau,n,h,t} = 1 \quad \forall k, \tau \in \mathcal{T}_k.
        \end{equation}
        
         \textbf{Precedence Constraints:} For any dependency $(\tau_i, \tau_j) \in \mathcal{E}_k$, the start time of $\tau_j$ must be after the finish time of $\tau_i$.
         
         \textbf{Deadline Satisfaction:} The completion time of the final sub-task of job $k$ must be no later than its deadline $D_k$.
         
         \textbf{Node Resource Capacity:} The total resource demand of tasks scheduled on node $n$ at time $t$ cannot exceed its capacity.
        \begin{equation}
        \sum_{k \in \mathcal{J}} \sum_{\tau \in \mathcal{T}_k} r_{\tau,h} x_{k,\tau,n,h,t} \le C_{n,h}^{comp} \quad \forall n,h\in h_\tau, t.
        \end{equation}

             \textbf{Migration Transmission Time Constraint:} For any subtask \(\tau\) migrated from node \(n_{src}\) to \(n_{dest}\) (i.e., \(x_{k,\tau,n_{src},h,t_1}=0\) and \(x_{k,\tau,n_{dest},h,t_2}=1\) with \(t_2 > t_1\)), the start time of \(\tau\) at \(n_{dest}\) must account for data migration time:
                   \begin{equation}
\begin{aligned}
start_{k,\tau,n_{dest},h,t_2} \geq &finish_{k,\tau_{pre},n_{src},h',t_1} + \\&t_{migrate,\tau,n_{src},n_{dest}}, \\ 
\end{aligned}
\end{equation}
 \begin{equation}
\forall (\tau_{pre}, \tau) \in E_k, h \in h_\tau, h' \in h_{\tau_{pre}}, 
\end{equation}
               where \(finish_{k,\tau_{pre},n_{src},h',t_1}\) is the completion time of predecessor subtask \(\tau_{pre}\) at \(n_{src}\), and \(t_{migrate,\tau,n_{src},n_{dest}}\) is calculated via Eq. (\ref{eq6}).

     \textbf{Link Bandwidth Constraint:} For each communication link \((n_{src}, n_{dest}) \in \mathcal{L}\), the total data transmission rate of all concurrent migration tasks must not exceed the link bandwidth:
      \begin{equation}
\begin{aligned}
 \sum_{\substack{
 migrate(\tau, n_{src} \to n_{dest})}} \frac{d_\tau}{t_{migrate,\tau,n_{src},n_{dest}}} \leq B_{n_{src},n_{dest}},
  \end{aligned}
\end{equation}
\begin{equation}
 \forall t, (n_{src}, n_{dest}) \in \mathcal{L},
\end{equation}
were \(migrate(\tau, n_{src} \to n_{dest})\) denotes that subtask \(\tau\) is migrated from \(n_{src}\) to \(n_{dest}\), and \(\tau'\) is the time interval of data transmission.
    
    \item \textbf{Coupling Constraint:} The total power consumption of CPN node $n$ co-located at bus $i$ is determined by the nonlinear power model of heterogeneous hardware and scheduling decisions:
     \begin{equation}
\begin{aligned}
    P_{i,n,t}^{CPN} = \sum_{h \in h_\tau} [ P_{n,h}^{idle} +  &(\alpha_{n,h} \cdot u_{n,h,t}^2 + \beta_{n,h} \cdot u_{n,h,t})\cdot \\  &(P_{n,h}^{peak} - P_{n,h}^{idle}) ].
   \end{aligned}
\end{equation}
The CPN power consumption is a nonlinear function, while the real-time economic dispatch (ED) of the power grid is a linear programming problem. The direct coupling of these two will lead to a significant increase in the computational complexity of the optimization problem. To balance accuracy and tractability, we can use piecewise linearization to handle the nonlinear terms \cite{chen2021data}.
    
    
    \item \textbf{Carbon Budget Constraint:} A long-term constraint on total carbon emissions is imposed to ensure sustainability goals are met.
    \begin{equation}
    \sum_{\omega \in \Omega} \pi_\omega \sum_{t=1}^{T} E_{t,\omega} \le E_{budget}.
    \end{equation}
    Due to its long-term nature, this constraint is difficult to handle directly in a short-term optimization. It will be managed implicitly through the design of the DRL agent's reward function, as detailed in the next section.

\end{itemize}

\section{Two-Stage Co-Optimization (TSCO) Framework}\label{sec-Iv}

\begin{figure}[!t]
   \centering
   \includegraphics[width=3 in]{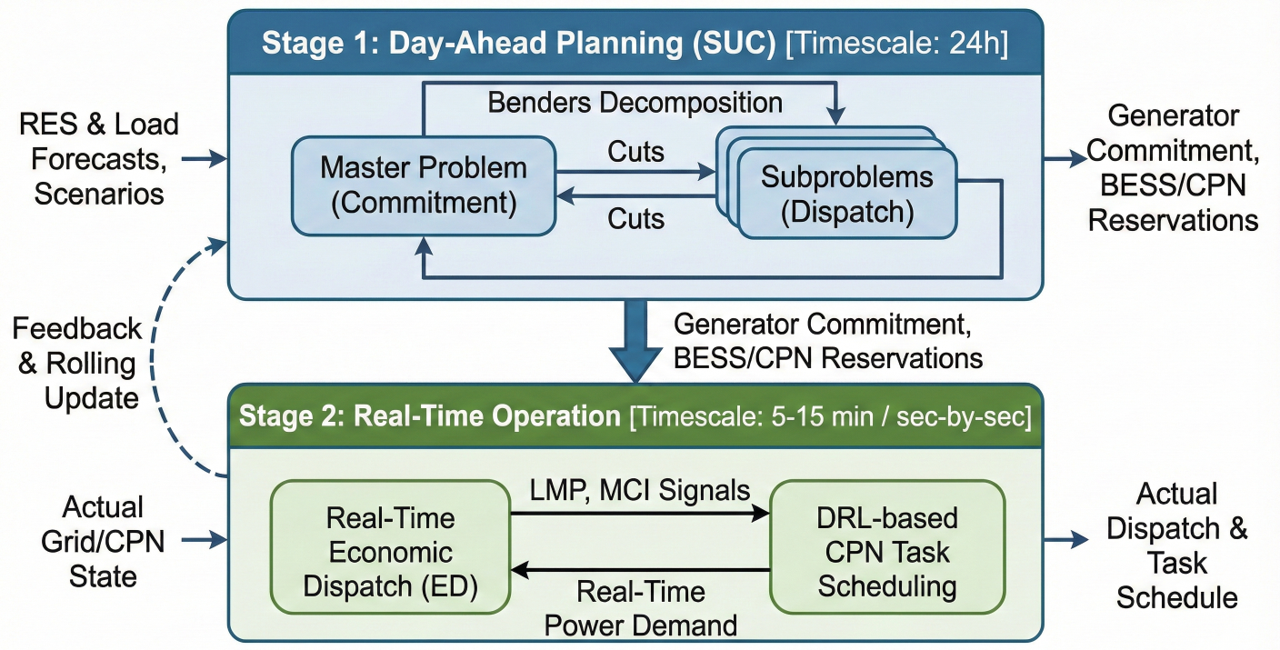}
   \caption{Two-Stage Co-Optimization (TSCO) framework for CPN and power grid collaborative optimization.}
   \label{fig1}
   \vspace{-0.5cm}
\end{figure}

The joint optimization problem formulated in Section \ref{sec-III} is a large-scale, non-convex, mixed-integer stochastic program, whose direct solution is computationally prohibitive at realistic scales \cite{tian2024distributed}. Our TSCO framework combines model-based optimization for long-term, system-wide planning with a model-free, AI-based approach for fast, adaptive real-time control, as shown in Fig. \ref{fig1}.

\subsection{Hierarchical Stochastic Optimization Structure}
The TSCO framework decomposes the problem into two distinct parts, reflecting the power system's natural hierarchy.
\begin{itemize}
    \item \textbf{Stage 1 (Day-Ahead Planning):} This stage solves an SUC problem for the upcoming 24-hour horizon. It makes the ``here-and-now" decisions, which are binding across all potential future scenarios. These decisions include the commitment (on/off) status $u_{g,t}$ of conventional generators and high-level energy reservation for BESS and CPN workload classes. The objective is to minimize the total expected cost over all scenarios, setting the operational envelope for the next stage. This stage is computationally intensive but is performed only once per day.
    \item \textbf{Stage 2 (Real-Time Operation):} This stage operates at a much faster timescale (e.g., 5-15 minute intervals) and makes ``wait-and-see" recourse decisions as uncertainty unfolds. As the actual RES generation and CPN task arrivals are revealed, this stage executes two parallel, tightly coupled processes:
    \begin{enumerate}
        \item \textbf{Real-Time ED:} Solves for the optimal power output of the committed generators and the charge/discharge schedule of BESS to meet the actual load at minimum cost, while respecting all grid constraints.
        \item \textbf{Real-Time CPN Task Scheduling:} A DRL agent makes granular, second-by-second scheduling decisions, assigning individual tasks to specific resources within the CPN nodes. Its decisions are informed by the real-time grid state, namely prices and carbon intensity, provided by the ED.
    \end{enumerate}
    \end{itemize}
    
     The interface between the two stages is designed to ensure tight coupling while maintaining computational tractability. Stage 1 outputs three types of binding constraints to Stage 2: (1) the commitment status \(u_{g,t}^*\) of conventional generators, which restricts the set of dispatchable units in real-time ED; (2) the energy reservation bounds \((SOC_{b,t}^{reserve, min}, SOC_{b,t}^{reserve, max})\) for BESS, ensuring sufficient flexibility to accommodate real-time RES volatility; (3) the per-node computational resource reservation \(C_{n,t}^{comp, reserve}\) for CPN, which limits the maximum resource utilization of each node to avoid overloading beyond the day-ahead planning envelope. 
     In return, Stage 2 feeds back key real-time operational data to Stage 1 for iterative improvement. It includes the deviation between actual and forecasted RES generation, the CPN task completion rate and deadline adherence, and the cumulative carbon emissions up to the current time. This feedback is used to refine the scenario set \(\Omega\) for the next day’s SUC problem.
     The day-ahead plan is updated once every 24 hours, while the real-time stage fetches the latest day-ahead constraints every 5-15 minutes and reports cumulative operational data at the end of each day.


\subsection{Benders Decomposition for the Day-Ahead SUC}
The day-ahead SUC problem is a large-scale MILP due to the combination of binary commitment variables and a large number of scenarios representing RES uncertainty. To solve it efficiently, we employ Benders decomposition \cite{pecci2025regularized}, a classic technique for problems with this structure. The method iteratively decomposes the problem into a simpler master problem and a set of independent subproblems.
\begin{itemize}
    \item \textbf{Master Problem:} The master problem determines the integer variables in the first stage, the unit commitment schedules $\{u_{g,t}\}$ for all $g \in \mathcal{G}_C$ over the horizon $T$. It is a pure integer program that minimizes the sum of startup/shutdown costs and an estimated future cost, $\theta$, which represents the expected operational cost from the subproblems.
    \begin{equation}
    \min \sum_{t=1}^{T} \sum_{g \in \mathcal{G}_C} (SU_g(u_{g,t}) + SD_g(u_{g,t})) + \theta.
    \end{equation}
    Subject to: Generator minimum up/down time constraints; Benders optimality and feasibility cuts (added iteratively).

    \item \textbf{Subproblems:} For a given commitment schedule $\{\bar{u}_{g,t}\}$ provided by the master problem, a separate, continuous linear program (LP) is solved for each scenario $\omega \in \Omega$. Each subproblem represents the economic dispatch problem for that scenario, minimizing the fuel and carbon costs subject to grid constraints.
    \begin{equation}
    \min \sum_{t=1}^{T} \left( \sum_{g \in \mathcal{G}_C} C_g(P_{g,t,\omega}) + \lambda_{CO2} E_{t,\omega} \right).
    \end{equation}
    Subject to: Power balance, line limits, generator output limits (for committed units), BESS constraints.
    
    \item \textbf{Algorithm Flow and Cut Generation:} The algorithm proceeds iteratively:
    \begin{enumerate}
        \item The master problem is solved to obtain a candidate commitment schedule.
        \item This schedule is passed to the subproblems, which are solved in parallel for all scenarios.
        \item If any subproblem is infeasible (i.e., the commitment schedule cannot satisfy the load), its dual rays are used to construct a \textbf{Benders feasibility cut}, which is added to the master problem to exclude this infeasible solution.
        \item If all subproblems are feasible, their optimal dual variable values are used to construct a \textbf{Benders optimality cut}. This cut is a linear inequality that provides a lower bound on the recourse cost $\theta$ and is added to the master problem.
        \item The process repeats until the lower bound from the master problem and the upper bound from the subproblems converge within a specified tolerance.
    \end{enumerate}
\end{itemize}

\subsection{DRL-based Real-Time CPN Task Scheduling}
While the SUC/ED provides an economically optimal power dispatch plan, it is far too slow for the dynamic, fine-grained scheduling required within the CPN. For this, we propose a DRL-based approach. A DRL agent can learn a complex scheduling policy through interaction with the environment \cite{liu2024dnn}, enabling it to make near-instantaneous decisions that are adaptive to both CPN and grid conditions.

\textbf{Markov Decision Process (MDP) Formulation:} The CPN scheduling problem is formulated as an MDP defined by the tuple $(\mathcal{S}, \mathcal{A}, \mathcal{P}, \mathcal{R}, \gamma)$:
    \begin{itemize}

        \item \textbf{State ($s_t \in \mathcal{S}$):} The state provides a comprehensive snapshot of the entire system at time $t$. It is a high-dimensional vector including:
        \begin{enumerate}
            \item \textit{CPN State:} Characteristics of tasks in the queue (e.g., resource requirements, deadlines), current resource and bandwidth utilization, and power consumption of each CPN node.
            \item \textit{Grid State:} Real-time RES generation levels, BESS state-of-charge, and crucially, the real-time locational marginal price (LMP) and marginal carbon intensity (MCI) for each bus hosting a CPN node. These signals are provided by the real-time ED solution.
       \end{enumerate}
            Furthermore, the state vector \(s_t\) is augmented with two day-ahead constraint tracking metrics:
         \begin{equation}
   Remaining_{n,h,t} = C_{n,h,t}^{comp, reserve}-\sum_{k,\tau} r_{\tau,h} x_{k,\tau,n,h,t-1},
    \end{equation}
         \begin{equation}
   Budget_t = E_{budget} - \sum_{\tau=1}^t \sum_{\omega \in \Omega} \pi_\omega E_{\tau,\omega},
    \end{equation}
The former remains a computational resource reserve of node $n$ at time $t$. The latter remains the carbon budget for the planning horizon. These metrics enable the DRL agent to proactively avoid approaching constraint boundaries.

        \item \textbf{Action ($a_t \in \mathcal{A}$):} For the task at the head of the queue, the agent selects an action from a discrete set. An action is a tuple $(n,h, \tau_{type})$ representing the decision to assign the task to node $n$ to be processed by resource type $\tau_{type}$ (e.g., CPU, GPU). The action space also includes deferring the task. The action space is further constrained by the day-ahead resource reservations from Stage 1. For any task assignment action \((n,h,\tau_{type})\), the total resource demand of tasks scheduled on node $n$ at time $t$, i.e., \(\sum_{k,\tau} r_{\tau,h} x_{k,\tau,n,h,t}\) must not exceed the day-ahead reserved capacity \(C_{n,h,t}^{comp, reserve}\). This constraint is enforced directly in the DRL agent’s action selection process by masking infeasible assignments, e.g., nodes with remaining capacity $\leq$ task resource requirement \(r_\tau,h\).
        \item \textbf{Reward ($R_t \in \mathcal{R}$):} The reward function is carefully designed to guide the agent towards the overall optimization objective. It is a weighted sum of multiple components:
   \begin{equation}
\begin{aligned}
        R_t = & w_{rev} \cdot \text{Revenue}_t - w_{cost} \cdot \text{Cost}_t\\& - w_{carb} \cdot \text{Carbon}_t - w_{pen} \cdot \text{Penalty}_t,
       \end{aligned}
\end{equation}
            where $\text{Revenue}_t$ is a positive reward for successfully completing a job; $\text{Cost}_t$ is the electricity cost of executing the scheduled task, calculated as $P_{n,t}^{CPN} \times \text{LMP}_{i,t}$; $\text{Carbon}_t$ denotes the carbon cost, calculated as $P_{n,t}^{CPN} \times \text{MCI}_{i,t}$. To enforce the long-term budget $E_{budget}$, this term is augmented using the Lyapunov optimization technique. A virtual carbon queue $Q_t$ is maintained, updating as 
            \begin{equation}
            Q_{t+1} = \max(0, Q_t + \text{Carbon}_t - E_{budget}/T).
            \end{equation}
            The reward is then penalized by an additional term proportional to $Q_t\text{Carbon}_t$, which strongly discourages carbon-intensive actions when the system is already over its carbon budget.
            And the $\text{Penalty}_t$ represents a large negative penalty for missing a task's deadline. The intuition behind this virtual carbon queue is to track the deviation between cumulative carbon emissions and the long-term budget. \(Q_t\) increases when the current carbon emission $\text{Carbon}_t$ exceeds the average allowable emission \(E_{budget}/T\) and resets to 0 if emissions are within the budget. When \(Q_t\) grows, indicating the system is approaching or exceeding the carbon budget, the additional penalty term $Q_t\text{Carbon}_t$ in the reward function strongly discourages the DRL agent from assigning tasks to high-carbon-intensity nodes. It then ensures that short-term scheduling decisions do not violate the long-term sustainability goal while avoiding excessive sacrifice of economic efficiency.

       \item \textbf{Constraint Adherence and Feedback Mechanism:} To address potential deviations from the day-ahead plan, a two-tier feedback mechanism is implemented:

       \begin{enumerate}
            \item \textit{Real-Time Infeasibility Correction:} If the real-time ED detects that the CPN power demand \(P_{i,t}^{CPN}\) violates grid constraints, such as transmission line limits or generator ramp constraints, the ED module sends a ``constraint violation signal” to the DRL agent. The agent then temporarily narrows its action space by increasing the penalty weight \(w_{pen}\) for tasks assigned to nodes causing violations, or masking those nodes for 1-2 time steps until the grid state stabilizes.
            \item \textit{Rolling Day-Ahead Plan Update:} If the cumulative deviation between real-time CPN resource utilization and day-ahead reserve, i.e., \(\sum_t |\sum_{k,\tau} r_\tau x_{k,\tau,n,t} - C_{n,t}^{comp, reserve}|\), exceeds a predefined threshold over 4 consecutive hours, the day-ahead SUC module is triggered to perform a mid-day rolling update. Then, the algorithm adjusts the remaining 24-hour reserve allocation and generator commitment status based on the real-time data.
       \end{enumerate}
         \end{itemize}
         
       Given the large, continuous state space and discrete action space, a value-based DRL algorithm such as Deep Q-Network (DQN) or its advanced variants (e.g., Dueling DQN, Rainbow) is suitable \cite{gok2024dynamic}. The agent's policy $\pi(a_t|s_t)$ is represented by a deep neural network that approximates the optimal action-value function $Q^*(s,a)$. The agent is trained offline on a rich dataset of historical system states and transitions, and then deployed for fast online inference.

\subsection{Overall TSCO Algorithm and Its Complexity Analysis}
The complete operational flow of the TSCO framework integrates the day-ahead planning and real-time control stages. The step-by-step procedure is outlined in Alg. \ref{alg:tsco}.

\begin{algorithm}[!t]
\caption{Two-Stage Co-Optimization (TSCO) Framework}\label{alg:tsco}
\SetKwInOut{Input}{Input}
\SetKwInOut{Output}{Output}

\Input{Set of RES/CPN scenarios $\Omega$ with probabilities $\pi_\omega$}
\Output{24-hour unit commitment $\{u_{g,t}^*\}$, Real-time power dispatch, Real-time CPN task schedule}
\BlankLine

\underline{Stage 1: Day-Ahead SUC (solved once daily):}\\
1. Initialize Benders master problem with generator constraints \\
2. \Repeat{lower and upper bounds converge}{
    3. Solve MILP master problem to get candidate commitment $\{\bar{u}_{g,t}\}$ \\
    4. \For{each scenario $\omega \in \Omega$ \KwSty{in parallel}}{
        5. Solve LP dispatch subproblem with fixed commitments $\{\bar{u}_{g,t}\}$ \\
        6. \eIf{subproblem is infeasible}{
            Generate and add a feasibility cut to the master problem
        }{
            Generate and add an optimality cut to the master problem
        }
    }
}
7. Obtain final 24-hour unit commitment schedule $\{u_{g,t}^*\}$ \\
\BlankLine

\underline{Stage 2: Real-Time Operation (for $t=1, \dots, T$):}\\
1. \For{$t=1, \dots, T$}{
    2. Observe actual RES generation $P_{g,t}^{R,actual}$ and new CPN job arrivals \\
    3. Update CPN task queue \\
    4. Solve real-time Economic Dispatch for $\{u_{g,t}^*\}$ and BESS \\
    5. Obtain real-time LMPs and MCIs for all CPN node locations \\
    6. Construct state vector $s_t \leftarrow$ (CPN state + Grid state + resource/carbon budget) \\
    7. Mask infeasible actions, DRL agent takes action $a_t \leftarrow \pi(s_t)$ to schedule task \\
    8. Update CPN power demand $P_{i,t}^{CPN}$ based on action $a_t$ \\
    9. Correct grid constraint violations (if any), execute dispatch/scheduling \\
    10. Update BESS SOC, CPN resource status and virtual carbon queue
}

\end{algorithm}

Additionally, the computational complexity of the TSCO framework is best analyzed by examining its two stages.



\subsubsection{Stage 1: Day-Ahead SUC}
The SUC problem is a mixed-integer programming (MIP) problem, which is NP-hard. Solving the full extensive form directly is computationally prohibitive for realistic system sizes and a large number of scenarios. Benders decomposition is employed to manage this complexity.
\begin{itemize}
    \item \textbf{Master Problem:} The master problem is a MILP. Its complexity is, in the worst case, exponential in the number of integer variables, which is proportional to the number of conventional generators and the length of the time horizon ($|\mathcal{G}_C| \times T$). The size of the master problem also grows with each iteration as Benders cuts are added.
    \item \textbf{Subproblems:} For each $\Omega$ scenario, a linear program (LP) is solved. The complexity of solving an LP with modern interior-point methods is polynomial in the number of variables and constraints. Since the subproblems are independent for a given commitment schedule, they can be solved in parallel. The time taken per iteration for this step is thus equivalent to solving a single LP.
   
\end{itemize}

\subsubsection{Stage 2: Real-Time Operation}
This stage must operate quickly at each time step $t$.
\begin{itemize}
    \item \textbf{Real-Time Economic Dispatch:} This is a standard LP, similar in structure to a Benders subproblem but for a single realized scenario. As an LP, it can be solved very efficiently in polynomial time, which is essential for real-time control.
    \item \textbf{DRL-based CPN Scheduling:} The online decision-making process involves a single forward pass through the trained deep neural network. The complexity of a forward pass is approximately $O(\sum_{l=1}^{L} N_l \times N_{l-1})$, where $L$ is the number of layers and $N_l$ is the number of neurons in layer $l$. This computation is extremely fast and independent of the complexity of the underlying system dynamics, making it highly suitable for real-time, low-latency scheduling decisions. Also, the computationally intensive training of the DRL agent is performed offline and does not impact the online operational complexity.
\end{itemize}

In summary, the TSCO framework strategically manages computational complexity by solving the NP-hard, large-scale planning problem (SUC) offline on a day-ahead basis, where longer computation times are acceptable. It then leverages highly efficient, polynomial-time algorithms (LP for ED) and fast neural network inference (DRL for scheduling) for the real-time operational stage, ensuring the framework is viable for practical deployment.

\section{Performance Evaluation}\label{sec-v}

A high-fidelity simulation environment is developed to assess its performance in terms of economic efficiency, environmental impact, grid stability, and CPN QoS. Additional experiments on convergence, scalability, ablation, and computational burden are conducted to further verify the framework’s engineering feasibility and core value.

\subsection{Simulation Setup}

The simulation framework is implemented in Python. The power system dynamics are modeled using PyPSA, a powerful open-source library for power system analysis. The CPN and the scheduling logic are implemented as a custom discrete-event simulator. The DRL agent is developed using PyTorch. For solving the MILP and LP problems in the Benders decomposition, we use the Gurobi optimizer.

The simulation is based on a modified IEEE 30-bus test system\footnote{https://icseg.iti.illinois.edu/ieee-30-bus-system/}. It includes 6 conventional thermal generators, 4 utility-scale BESS units, and 5 large-scale renewable generation sites (3 solar, 2 wind).To ensure realism, we use real-world time-series data to model the stochastic RES generation. Solar irradiance data for locations in California is sourced from the National Renewable Energy Laboratory's (NREL) National Solar Radiation Database (NSRDB)\footnote{https://catalog.data.gov/dataset/national-solar-radiation-database-nsrdb}. Wind power generation profiles for locations in Germany are obtained from the ENTSO-E Transparency Platform\footnote{https://www.entsoe.eu/data/power-stats/}. These datasets are used to generate 100 distinct 24-hour scenarios for the SUC problem.  


The CPN consists of 5 geo-distributed nodes, each co-located with a major load bus in the IEEE 30-bus system. The arrival patterns and resource requirements of computational jobs are derived from processed Google Cluster Data traces\footnote{https://www.kaggle.com/datasets/derrickmwiti/google-2019-cluster-sample}, which provide a realistic representation of large-scale data center workloads. The precedence constraints and dependency structures within jobs are modeled based on common scientific workflow patterns, such as pipeline workflow, available from the Pegasus Workflow Management System\footnote{https://pegasus.isi.edu/documentation/examples/}. 

To ground our environmental and economic calculations in reality, we incorporate two external data sources. Real-time marginal carbon intensity (MCI) data is obtained via the WattTime API, which provides 5-minute resolution data on the emissions impact of consuming an additional MWh of electricity in various grid regions\footnote{https://watttime.org/data-science/data-signals/marginal-co2/}. Historical hourly locational marginal price (LMP) data is sourced from the California Independent System Operator (CAISO) public database\footnote{https://www.gridstatus.io/live/caiso}. 
To ensure the generality of the experimental results, all values are presented as the mean ± standard deviation derived from 10 independent runs.

\begin{figure*}[!t]
   \centering
   \includegraphics[width=6.5 in]{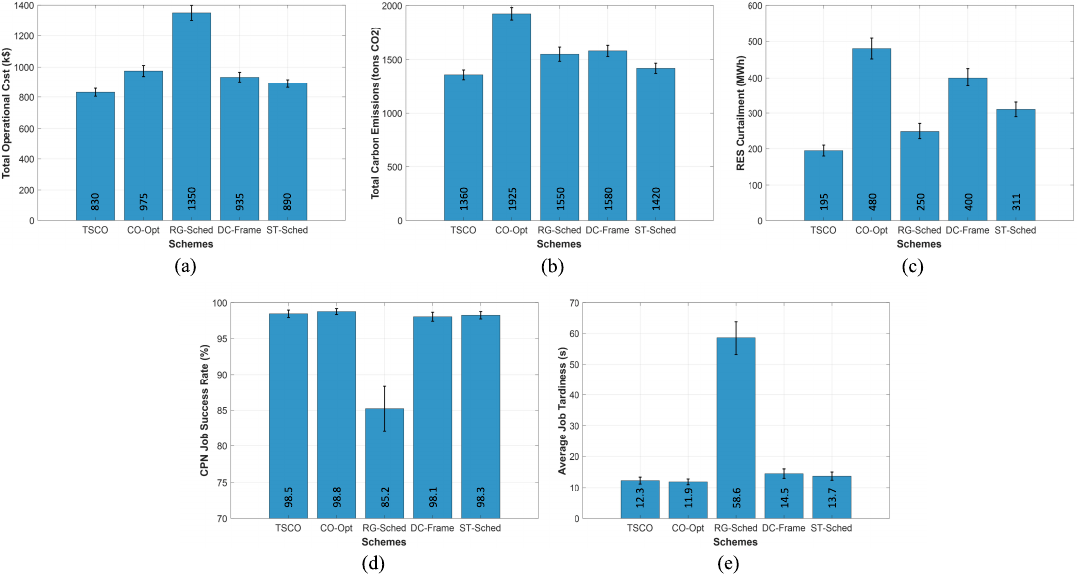}
   \caption{Baseline performance comparison. (a) Total operational cost; (b) Total carbon emissions; (c) RES curtailment; (d) CPN job success rate; (d) Average job tardiness.}
   \label{fig2}
   \vspace{-0.5cm}
\end{figure*}

\subsection{Comparison Schemes}

To rigorously evaluate the performance of our TSCO framework, we compare it against three baseline methods that represent alternative approaches to the problem:

\begin{itemize}
    \item \textbf{Cost-Only Optimizer (CO-Opt) \cite{li2024assessment}:} This baseline uses the same two-stage optimization architecture as TSCO but with the carbon price set to zero. It represents the current industry-standard approach of economic dispatch.
    \item \textbf{Renewable-Greedy Scheduler (RG-Sched) \cite{davoudi2025non}:} This is a heuristic-based CPN scheduling approach where tasks are always dispatched to the CPN node with the highest instantaneous RES power availability. 
    \item \textbf{Decoupled Framework (DC-Frame) \cite{zhong2024joint}:} It is the state-of-the-art in carbon-aware computing, where the power system and CPN are optimized separately. The power system operation is optimized first to generate a fixed 24-hour profile of electricity prices and carbon intensities. Subsequently, the CPN scheduler optimizes its task scheduling based on these static, pre-computed signals. 
    \item \textbf{DRL-based Spatiotemporal Scheduler (ST-Sched) \cite{wen2026green}, \cite{wen2024spatiotemporal}:} This baseline adopts a customized DRL approach for intra-CPN fine-grained scheduling. It features a triple-selection action space to optimize energy consumption, carbon emissions, and task delay while ensuring load balancing. 
\end{itemize}

\subsection{Comparison Performances}

\subsubsection{Baseline Performance Comparison}

In this scenario, all five methods were simulated over one week (168 hours) with the carbon price set at a representative value of $\$50$/ton.  The mean and standard deviation are present in Fig. \ref{fig2}.

TSCO simultaneously achieves the lowest total operational cost and the lowest carbon emissions. The CO-Opt baseline, being carbon-agnostic, minimizes only direct fuel costs, resulting in $41.5\%$ higher emissions and $17.5\%$ higher total costs once the carbon price is factored in. This is because it relies heavily on the cheapest available thermal generators, regardless of their carbon intensity. The RG-Sched heuristic, while intuitive, performs poorly across the board. By myopically chasing renewables, it ignores grid congestion and the economic cost of dispatching thermal generators to support its decisions, leading to the highest operational cost and only modest emission reductions. The DC-Frame performs better than the naive baselines but is still significantly outperformed by TSCO. Its reliance on static, day-ahead signals prevents it from adapting to real-time deviations between forecasted and actual grid conditions, leading to $12.7\%$ higher costs and $16.2\%$ higher emissions. The ST-Sched focuses on intra-CPN fine-grained scheduling via customized DRL. It demonstrates competitive performance relative to DC-Frame and CO-Opt but falls short of TSCO. It achieves $8.4\%$ higher operational costs and $4.4\%$ higher carbon emissions than TSCO. Its lack of cross-domain coordination with the power system prevents it from leveraging real-time grid dynamics (e.g., surplus renewable energy) to optimize overall efficiency.
A key finding is TSCO's ability to significantly reduce RES curtailment. By treating the CPN as a flexible load that can absorb surplus renewable generation in real-time, TSCO reduces curtailment by over $60\%$ compared to DC-Frame and CO-Opt, and by $37.1\%$ compared to ST-Sched. This demonstrates the value of co-optimization in turning a major energy consumer into a valuable grid-stabilizing asset.

Additionally, the economic and environmental gains from TSCO do not come at the expense of computational performance. TSCO maintains a high job success rate $98.5\%$ and low average tardiness $12.3$ s, nearly on par with the cost-only optimizer. This is a direct result of the DRL agent's reward function, which is designed to penalize deadline violations, forcing it to learn a policy that balances sustainability goals with QoS requirements. In contrast, the RG-Sched baseline suffers from a poor job success rate $85.2\%$ and high tardiness because its singular focus on RES availability often leads it to schedule tasks on nodes that are already congested, highlighting the need for a holistic system view.

\begin{figure*}[!t]
   \centering
   \includegraphics[width=6.5 in]{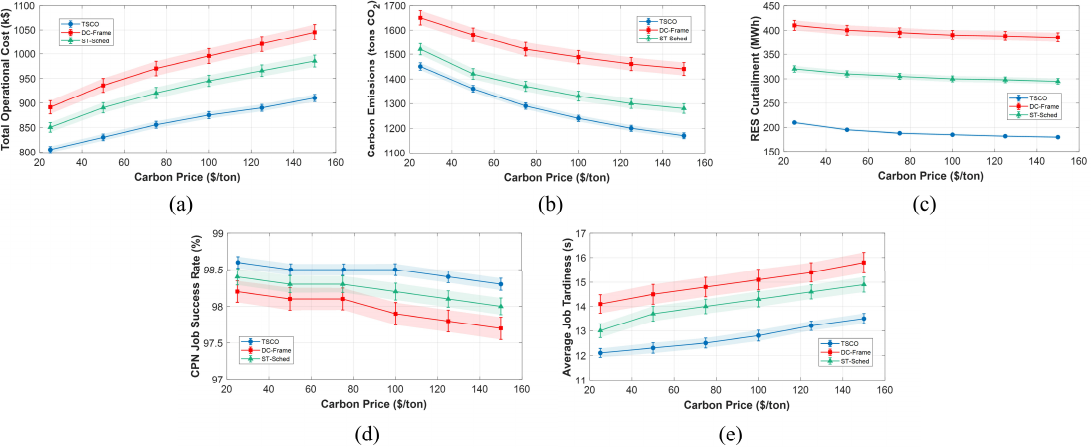}
   \caption{Sensitivity analysis with varying carbon price. (a) Total operational cost; (b) Total carbon emissions; (c) RES curtailment; (d) CPN job success rate; (d) Average job tardiness.}
   \label{fig3}
   \vspace{-0.2cm}
\end{figure*}

\subsubsection{Sensitivity to Carbon Price}

To analyze the trade-off between economic and environmental objectives, we varied the carbon price $\lambda_{CO_2}$ from $25$ to $150$/ton. This analysis focuses on the TSCO, DC-Frame, and ST-Sched methods, as the other two baselines are insensitive to carbon price by design. The comprehensive results are presented in Fig. \ref{fig3}.

TSCO achieves a superior economic-environmental trade-off across all metrics at every carbon price point. All three carbon-aware methods cut emissions with rising carbon prices, but TSCO does so with the highest efficiency at $19.3\%$. Economically, TSCO also demonstrates optimal performance. Its operational cost rises by the smallest margin with increasing carbon prices, remaining approximately $7\%$ lower than ST-Sched and $14\%$ lower than DC-Frame across all price points. ST-Sched’s single-domain intra-CPN DRL optimization delivers moderate cost and emission improvements over DC-Frame but falls short of TSCO. It lacks cross-domain coordination to leverage grid-side low-carbon, low-cost resources. DC-Frame’s decoupled design, reliant on static day-ahead grid signals, fails to adapt to real-time deviations. It leads to the highest operational costs and the least efficient emission reduction among them.
For RES curtailment, rising carbon prices drive greater renewable utilization for all methods, with TSCO recording the largest reduction at $14.3\%$. Its curtailment levels are $37.1\%$ lower than ST-Sched and less than half that of DC-Frame. This underscores TSCO’s unique ability to treat the CPN as a flexible load to absorb surplus grid-side renewable energy.

\begin{table*}[!t]
\centering
\caption{Convergence Performance Comparison}
\label{tab:convergence_perf}

\renewcommand{\arrayrulewidth}{0.8pt} 
\renewcommand{\tabcolsep}{10pt} 

{\fontsize{8}{10}\selectfont 
\begin{tabular}{m{2cm}||m{2cm}|m{3cm}|m{3cm}|m{3cm}} 
   \hline
  \textbf{Method} & \textbf{Load Intensity} & \textbf{Convergence Iterations} & \textbf{Convergence Time (min)} & \textbf{Bound Gap (\$)} \\
  \hline
  \multirow{3}*{Direct-MILP}     & Low    & - (infeasible) & Larger than 360 (timeout) & -             \\
  \cline{2-5}
  ~ & Medium & - (infeasible) & Larger than 360 (timeout) & -             \\
  \cline{2-5}
  ~ & High   & - (infeasible) & Larger than 360 (timeout) & -             \\
  \hline
  \multirow{3}*{Improved-Benders} & Low    & 28.3±2.1       & 45.7±3.2       & 128.5±15.3    \\
  \cline{2-5}
  ~ & Medium & 35.6±2.8       & 58.2±4.1       & 156.3±18.7    \\
  \cline{2-5}
    ~ & High   & 42.1±3.5       & 72.4±5.6       & 189.7±21.2    \\
    \hline
    \multirow{3}*{TSCO-Benders}    & Low    & 21.5±1.8       & 32.4±2.5       & 98.6±12.4     \\
    \cline{2-5}
    ~ & Medium & 27.8±2.3       & 43.1±3.7       & 121.4±16.5    \\
    \cline{2-5}
    ~ & High   & 33.2±2.9       & 55.8±4.8       & 145.2±19.3    \\
    \hline
\end{tabular}}

\vspace{1mm}
{\fontsize{8}{10}\selectfont }
   \vspace{-0.5cm}
\end{table*}

Notably, TSCO’s environmental and economic gains do not compromise CPN service quality. It retains the highest job success rate and lowest average tardiness across all carbon prices. ST-Sched also delivers strong QoS performance due to its refined intra-CPN load balancing. While DC-Frame experiences noticeable QoS degradation with rising carbon prices due to its rigid precomputed grid signals failing to account for real-time CPN congestion.

\subsection{Convergence Comparison}
To verify the efficiency and stability of the Benders decomposition in the day-ahead SUC stage, we compare two methods under different load intensities, including Direct-MILP without decomposition and Improved-Benders \cite{pecci2025regularized}. The results are summarized in Table \ref{tab:convergence_perf}. Among them, the mid-range workload represents an average daily task arrival rate of $800$ jobs per day with a total computational workload of $5\times10^{15}$ FLOPs, and an average resource requirement of $4$ CPU cores and $2$ GPU cores or equivalent TPU resources per job. The low and high loads are $0.8$ times and $1.2$ times the medium load, respectively.

Direct-MILP fails to converge within 6 hours for all scenarios due to the high dimensionality of the SUC problem. Both Benders-based methods achieve convergence, but TSCO-Benders outperforms Improved-Benders by $24.2\%-27.9\%$ in convergence iterations and $29.1\%-31.3\%$ in convergence time. This is because the tailored Benders cuts in TSCO effectively reduce the search space of the master problem. The bound gap of TSCO-Benders is consistently below $\$150$, indicating high solution accuracy. For medium load, TSCO-Benders takes an average of $43.1$ minutes to converge, which is acceptable for day-ahead planning.

\subsection{Scalability Verification}
We also expand the system scale in three dimensions to test the scalability of the TSCO framework. The results are shown in Table \ref{tab:scalability_perf}.

\begin{table*}[!t]
\centering
\caption{Scalability Performance Under Different System Scales}
\label{tab:scalability_perf}

\renewcommand{\arrayrulewidth}{0.8pt} 
\renewcommand{\tabcolsep}{10pt} 

{\fontsize{8}{10}\selectfont 
\begin{tabular}{m{0.6cm}||m{1.2cm}|m{1cm}|m{1.7cm}|m{1.7cm}|m{1.7cm}|m{2cm}|m{2.2cm}} 
    \hline
    \textbf{CPN Nodes} & \textbf{RES Scenarios} & \textbf{Gen. Count} & \textbf{Day-Ahead Conv. Time (min)} & \textbf{Day-Ahead Conv. Iterations} & \textbf{Real-Time ED Time (s/step)} & \textbf{DRL Inference Time (ms/task)} & \textbf{Performance Retention Rate (\%)} \\
    \hline
    5  & 100 & 6  & 43.1±3.7  & 27.8±2.3  & 1.2±0.1  & 8.5±0.6  & 100.0±0.0 \\
    \hline
    10 & 200 & 8  & 68.5±4.9  & 35.2±2.7  & 1.8±0.2  & 10.3±0.8 & 96.7±0.8  \\
    \hline
    15 & 300 & 10 & 95.3±6.2  & 42.7±3.1  & 2.5±0.3  & 12.1±1.0 & 93.5±1.1  \\
    \hline
    20 & 500 & 12 & 132.7±7.8 & 51.4±3.6  & 3.3±0.4  & 14.8±1.2 & 90.2±1.5  \\
    \hline
\end{tabular}}
\vspace{1mm}
{\fontsize{8}{10}\selectfont }
\end{table*}

\begin{figure*}[!t]
   \centering
   \includegraphics[width=6.5 in]{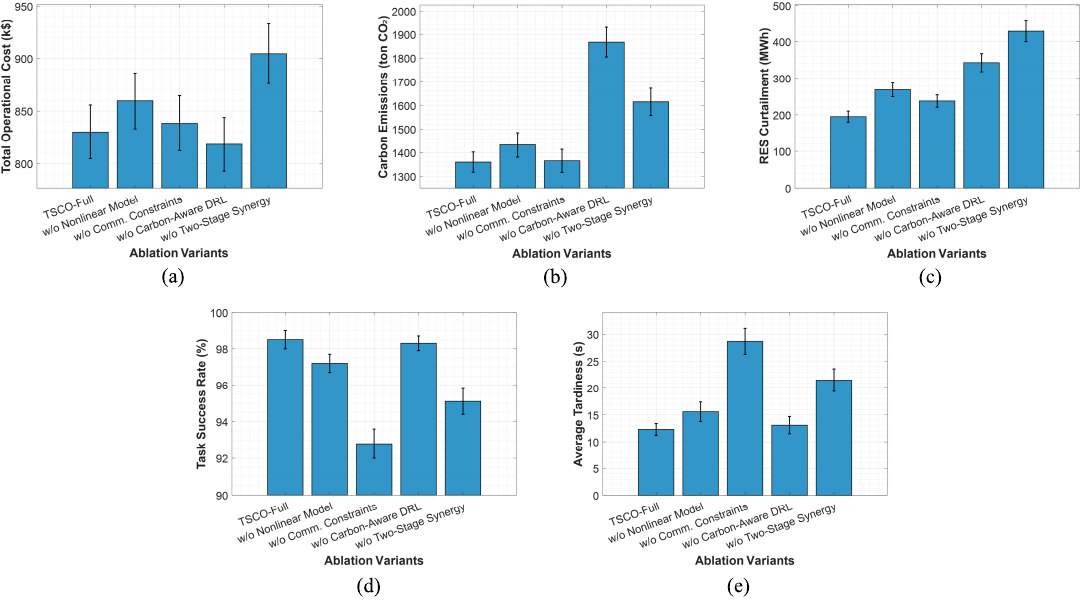}
   \caption{Ablation test. (a) Total operational cost; (b) Total carbon emissions; (c) RES curtailment; (d) CPN job success rate; (d) Average job tardiness.}
   \label{fig a}
   \vspace{-0.5cm}
\end{figure*}
As the system scale expands with 20 nodes, 500 scenarios, and 12 generators, the day-ahead convergence time increases to 132.7 minutes. It is still within the 24-hour planning window, and the real-time ED/DRL time remains at the millisecond-to-second level. The performance retention rate is $90.2\%$, meaning the carbon emission reduction rate and RES utilization only decrease by $9.8\%$ compared to the baseline. This demonstrates that TSCO’s hierarchical structure effectively isolates computational complexity. The day-ahead stage tolerates longer computation for large-scale optimization, while the real-time stage maintains fast response via LP and DRL inference. The framework thus meets the requirements of large-scale CPN-grid integration.

\subsection{Ablation Test}
To validate the necessity of key modules in the TSCO framework, we design four variants by removing core components one by one. In this test, we remove the nonlinear power consumption model of CPN from the TSCO and replace it with a simple linear model; Constraints of the communication link; Carbon-aware DRL; and integration with the two-stage process in TSCO. Its simulation parameters are set in accordance with those of Section \ref{sec-v}-C-1).

The results are presented in Fig. \ref{fig a}. Specifically, the nonlinear power model reduces carbon emissions by $7.1\%$ and operational cost by $3.0\%$ compared to the linear model, as it accurately captures GPU/TPU power dynamics and avoids over/underestimating CPN load. Meanwhile, communication constraints improve task success rate by $5.7\%$ and reduce average tardiness by $57.1\%$, as they prevent migration failures caused by bandwidth/latency bottlenecks. Moreover, the carbon-aware DRL reduces emissions by $28.3\%$ without significant economic loss, proving its ability to balance sustainability and cost. Furthermore, Two-stage synergy reduces operational cost by $8.4\%$ and RES curtailment rate by $55.6\%$, highlighting the value of real-time feedback to day-ahead planning.
Therefore, we integrate the full framework of these modules to achieve optimal comprehensive performance.

\subsection{Analysis of Computational Burden}

To verify the feasibility of the TSCO framework, we quantify the computational time of each stage under a standard server environment (CPU: Intel Xeon 8375C 3.0GHz; GPU: NVIDIA A100 40GB; Memory: 256GB). The results are shown in Table \ref{tab:computational_burden}.

\begin{table}[!t]
\centering
\caption{Computational Burden Data}
\label{tab:computational_burden}

\renewcommand{\arrayrulewidth}{0.8pt} 
\renewcommand{\tabcolsep}{10pt} 

{\fontsize{8}{10}\selectfont 
\begin{tabular}{m{2.2cm}||m{2.8cm}|m{2cm}} 
    \hline
    \textbf{Stage} & \textbf{Metric} & \textbf{Value (mins)} \\
    \hline
    \multirow{8}*{Day-Ahead SUC} & Average Convergence Iterations & 27.8±2.3 \\
    \cline{2-3}
    ~ & Average Time per Iteration (min) & 1.55±0.12 \\
    \cline{2-3}
    ~ & Average Total Convergence Time (min) & 43.1±3.7 \\
    \cline{2-3}
    ~ & Master Problem Solving Time (min/iter) & 0.32±0.05 \\
    \cline{2-3}
    ~ & Subproblems Solving Time (min/iter, parallel) & 1.23±0.10 \\
    \hline
    \multirow{3}*{Real-Time Operation} & ED Average Solving Time (s/step) & 1.2±0.1 \\
    \cline{2-3}
    ~ & DRL Inference Time (ms/task) & 8.5±0.6 \\
    \hline
    DRL Training& Total Training Time (hours) & 48.3±3.2 \\
    \hline
\end{tabular}}

\vspace{1mm}
{\fontsize{8}{10}\selectfont} 
   \vspace{-0.5cm}
\end{table}

The day-ahead SUC stage has an average convergence time of $43.1$ minutes, which is far less than the $24$-hour planning cycle, leaving sufficient time for scenario updates and human review. The real-time ED solves in $1.2$ seconds per step, namely a $5$-minute interval. Meanwhile, DRL infers in $8.5$ milliseconds per task, meeting the requirements of real-time scheduling with second-level response. The offline DRL training takes $48.3$ hours, which is acceptable as it only needs to be performed once before deployment with incremental fine-tuning every $3$ months based on new data. These quantitative results confirm that the TSCO framework balances optimization accuracy and computational efficiency, making it suitable for practical engineering deployment.

\section{Conclusion}\label{sec-vi}

This paper addresses the intertwined challenges of growing energy consumption in CPNs and power grid instability caused by high RES penetration. These two issues are critical for service computing sustainability and QoS.
To overcome these limitations, we propose a novel TSCO framework that synergistically coordinates CPN task scheduling and power system dispatch. By decomposing the complex problem into a day-ahead stochastic unit commitment stage and a real-time operational stage, the framework aligns service QoS requirements with grid sustainability goals.
Extensive simulations validate the TSCO framework’s superiority over baselines. It achieves a $16.2\%$ cut in carbon emissions and a $12.7\%$ reduction in operational costs, while cutting RES curtailment by more than $60\%$. Notably, these environmental and economic gains are achieved without compromising service reliability. The framework sustains a $98.5\%$ task success rate and limits average task tardiness to $12.3$s. This work advances cross-domain service optimization in CPNs, laying the groundwork for efficient, low-carbon service computing.


\bibliographystyle{IEEEtran}
\bibliography{IEEEabrv,mylib}

\begin{thebibliography}{10}
\providecommand{\url}[1]{#1}
\csname url@samestyle\endcsname
\providecommand{\newblock}{\relax}
\providecommand{\bibinfo}[2]{#2}
\providecommand{\BIBentrySTDinterwordspacing}{\spaceskip=0pt\relax}
\providecommand{\BIBentryALTinterwordstretchfactor}{4}
\providecommand{\BIBentryALTinterwordspacing}{\spaceskip=\fontdimen2\font plus
\BIBentryALTinterwordstretchfactor\fontdimen3\font minus \fontdimen4\font\relax}
\providecommand{\BIBforeignlanguage}[2]{{%
\expandafter\ifx\csname l@#1\endcsname\relax
\typeout{** WARNING: IEEEtran.bst: No hyphenation pattern has been}%
\typeout{** loaded for the language `#1'. Using the pattern for}%
\typeout{** the default language instead.}%
\else
\language=\csname l@#1\endcsname
\fi
#2}}
\providecommand{\BIBdecl}{\relax}
\BIBdecl

\bibitem{luo2025toward}
H.~Luo \emph{et~al.}, ``Toward edge general intelligence with multiple-large language model (multi-llm): Architecture, trust, and orchestration,'' \emph{IEEE Transactions on Cognitive Communications and Networking}, 2025.

\bibitem{yukun2024computing}
S.~Yukun \emph{et~al.}, ``Computing power network: A survey,'' \emph{China Communications}, vol.~21, no.~9, pp. 109--145, 2024.

\bibitem{liu2025joint}
J.~Liu \emph{et~al.}, ``Joint task coding and transfer optimization for edge computing power networks,'' \emph{IEEE Transactions on Network Science and Engineering}, 2025.

\bibitem{impram2020challenges}
S.~Impram \emph{et~al.}, ``Challenges of renewable energy penetration on power system flexibility: A survey,'' \emph{Energy Strategy Reviews}, vol.~31, p. 100539, 2020.

\bibitem{chen2025data}
S.~Chen, ``Data centres will use twice as much energy by 2030-driven by ai,'' \emph{Nature}, 2025.

\bibitem{teng2024privacy}
F.~Teng \emph{et~al.}, ``A privacy-preserving distributed economic dispatch method for integrated port microgrid and computing power network,'' \emph{IEEE Transactions on Industrial Informatics}, vol.~20, no.~8, pp. 10\,103--10\,112, 2024.

\bibitem{wang2025providing}
Y.~Wang, Q.~Guo, and M.~Chen, ``Providing load flexibility by reshaping power profiles of large language model workloads,'' \emph{Advances in Applied Energy}, p. 100232, 2025.

\bibitem{luo2024bc4llm}
H.~Luo, J.~Luo, and A.~V. Vasilakos, ``Bc4llm: A perspective of trusted artificial intelligence when blockchain meets large language models,'' \emph{Neurocomputing}, vol. 599, p. 128089, 2024.

\bibitem{jiang2025bargaining}
Z.~Jiang and Y.~Guo, ``Bargaining-based approach for dynamic operating envelope allocation in distribution networks,'' \emph{IEEE Transactions on Smart Grid}, 2025.

\bibitem{yang2024secure}
S.~Yang \emph{et~al.}, ``Secure distributed control for demand response in power systems against deception cyber-attacks with arbitrary patterns,'' \emph{IEEE Transactions on Power Systems}, vol.~39, no.~6, pp. 7277--7290, 2024.

\bibitem{piontek2023carbon}
T.~Piontek, K.~Haghshenas, and M.~Aiello, ``Carbon emission-aware job scheduling for kubernetes deployments,'' \emph{Journal of supercomputing}, vol.~80, pp. 549--569, 2023.

\bibitem{xie2025priority}
R.~Xie \emph{et~al.}, ``Priority-aware task scheduling in computing power network-enabled edge computing systems,'' \emph{IEEE Transactions on Network Science and Engineering}, 2025.

\bibitem{ma2025optimizing}
B.~Ma \emph{et~al.}, ``Optimizing profit and delay in computing power network via deep deterministic policy gradient: A task decomposition and computing path optimization approach,'' \emph{IEEE Transactions on Services Computing}, 2025.

\bibitem{chen2024two}
Q.~Chen \emph{et~al.}, ``Two-stage evolutionary search for efficient task offloading in edge computing power networks,'' \emph{IEEE Internet of Things Journal}, vol.~11, no.~19, pp. 30\,787--30\,799, 2024.

\bibitem{wen2026green}
W.~Wen \emph{et~al.}, ``Green orchestra: Joint spatiotemporal task scheduling and hybrid energy coordination in computing power networks,'' \emph{IEEE Transactions on Cognitive Communications and Networking}, 2026.

\bibitem{wen2024spatiotemporal}
------, ``Spatiotemporal task scheduling for green computing in computing power networks,'' in \emph{2024 IEEE Global Communications Conference}.\hskip 1em plus 0.5em minus 0.4em\relax IEEE, 2024, pp. 2449--2454.

\bibitem{zhong2024joint}
W.~Zhong \emph{et~al.}, ``Joint energy-computation management for electric vehicles under coordination of power distribution networks and computing power networks,'' \emph{IEEE Transactions on Smart Grid}, vol.~16, no.~2, pp. 1549--1561, 2025.

\bibitem{luo2024multi}
H.~Luo \emph{et~al.}, ``A multi-chain consensus for power big data transaction in generation-grid-load-storage integrated networks,'' in \emph{2024 IEEE Global Communications Conference}.\hskip 1em plus 0.5em minus 0.4em\relax IEEE, 2024, pp. 2455--2460.

\bibitem{ma2024study}
T.~Ma \emph{et~al.}, ``Study on multi-time scale frequency hierarchical control method and dynamic response characteristics of the generation-grid-load-storage type integrated system under double-side randomization conditions,'' \emph{Applied Energy}, vol. 367, p. 123436, 2024.

\bibitem{chowdhury2025optimal}
M.~Chowdhury \emph{et~al.}, ``Optimal power flow (opf) analysis for ac--dc active distribution networks utilizing second-order cone programming (socp) approach,'' \emph{IEEE Transactions on Industrial Informatics}, 2025.

\bibitem{zhang2025unlocking}
Y.~Zhang \emph{et~al.}, ``Unlocking the flexibilities of data centers for smart grid services: Optimal dispatch and design of energy storage systems under progressive loading,'' \emph{Energy}, vol. 316, p. 134511, 2025.

\bibitem{ye2024deep}
Z.~Ye \emph{et~al.}, ``Deep learning workload scheduling in gpu datacenters: A survey,'' \emph{ACM Computing Surveys}, vol.~56, no.~6, pp. 1--38, 2024.

\bibitem{ma2025greening}
H.~Ma \emph{et~al.}, ``Greening edge ai: Optimizing inference accuracy and reducing carbon emissions with renewable energy,'' \emph{IEEE Internet of Things Journal}, 2025.

\bibitem{xu2024optimal}
J.~Xu \emph{et~al.}, ``Optimal task scheduling and resource allocation for self-powered sensors in internet of things: An energy efficient approach,'' \emph{IEEE Transactions on Network and Service Management}, vol.~21, no.~4, pp. 4410--4420, 2024.

\bibitem{chen2024drdst}
R.~Chen \emph{et~al.}, ``Drdst: Low-latency dag consensus through robust dynamic sharding and tree-broadcasting for iov,'' \emph{IEEE Transactions on Mobile Computing}, 2025.

\bibitem{xie2017energy}
G.~Xie \emph{et~al.}, ``Energy-aware processor merging algorithms for deadline constrained parallel applications in heterogeneous cloud computing,'' \emph{IEEE Transactions on Sustainable Computing}, vol.~2, no.~2, pp. 62--75, 2017.

\bibitem{rengel2023optimal}
A.~Rengel \emph{et~al.}, ``Optimal insertion of energy storage systems considering the economic dispatch and the minimization of energy not supplied,'' \emph{Energies}, vol.~16, no.~6, p. 2593, 2023.

\bibitem{maroufi2025power}
S.~M. Maroufi \emph{et~al.}, ``Power management of hybrid flywheel-battery energy storage systems considering the state of charge and power ramp rate,'' \emph{IEEE Transactions on Power Electronics}, 2025.

\bibitem{zou2024aggregator}
Y.~Zou, Y.~Xu, and J.~Li, ``Aggregator-network coordinated peer-to-peer multi-energy trading via adaptive robust stochastic optimization,'' \emph{IEEE Transactions on Power Systems}, vol.~39, no.~6, pp. 7124--7137, 2024.

\bibitem{bouhouras2024congestion}
A.~S. Bouhouras \emph{et~al.}, ``Congestion management in coupled tso and dso networks,'' \emph{Electric Power Systems Research}, vol. 229, p. 110145, 2024.

\bibitem{chen2021data}
J.~Chen, W.~Wu, and L.~A. Roald, ``Data-driven piecewise linearization for distribution three-phase stochastic power flow,'' \emph{IEEE Transactions on Smart Grid}, vol.~13, no.~2, pp. 1035--1048, 2021.

\bibitem{tian2024distributed}
F.~Tian, H.~Liu, and W.~Yu, ``A distributed decomposition algorithm for solving large-scale mixed integer programming problem,'' \emph{Science China Information Sciences}, vol.~67, no.~12, p. 222205, 2024.

\bibitem{pecci2025regularized}
F.~Pecci and J.~D. Jenkins, ``Regularized benders decomposition for high performance capacity expansion models,'' \emph{IEEE Transactions on Power Systems}, 2025.

\bibitem{liu2024dnn}
Z.~Liu \emph{et~al.}, ``Dnn partitioning, task offloading, and resource allocation in dynamic vehicular networks: A lyapunov-guided diffusion-based reinforcement learning approach,'' \emph{IEEE Transactions on Mobile Computing}, vol.~24, no.~3, pp. 1945--1962, 2025.

\bibitem{gok2024dynamic}
M.~G{\"o}k, ``Dynamic path planning via dueling double deep q-network (d3qn) with prioritized experience replay,'' \emph{Applied Soft Computing}, vol. 158, p. 111503, 2024.

\bibitem{li2024assessment}
J.~Li \emph{et~al.}, ``An assessment methodology for the flexibility capacity of new power system based on two-stage robust optimization,'' \emph{Applied Energy}, vol. 376, p. 124291, 2024.

\bibitem{davoudi2025non}
M.~Davoudi, M.~Chen, and J.~Qin, ``Non-preemptive scheduling of flexible loads in smart grids via convex optimization,'' \emph{IEEE Transactions on Control of Network Systems}, 2025.

\end{thebibliography}

\vfill

\end{document}